\documentclass[10pt,journal]{IEEEtran}

\usepackage{graphicx,times,cite,amsmath, amssymb, amsthm, mathrsfs, epsfig, array, subfigure}

\usepackage{multirow}

\usepackage[colorlinks, linkcolor=black, anchorcolor=black, citecolor=black, urlcolor=black]{hyperref}
\usepackage{cite}

\usepackage{color}

\allowdisplaybreaks[4]
\newcommand{\figwidthT}{2.5in}
\newcommand{\figwidthS}{2.8in}
\newcommand{\figwidthM}{2.8in}

\newcommand{\figwidthW}{0.83\linewidth}
\newcommand{\fref}[1]{Fig.~\ref{#1}}

\usepackage[acronym]{glossaries} 
\newacronym{iid}{i.i.d.}{independent and identically distributed}
\newacronym{SNR}{SNR}{signal-to-noise ratio}
\newacronym{SINR}{SINR}{signal-to-interference-plus-noise ratio}
\newacronym{AWGN}{AWGN}{additive white Gaussian noise}
\newacronym{MAC}{MAC}{multiple access channel}
\newacronym{BF}{BF}{bit-flipping}
\newacronym{IoT}{IoT}{Internet-of-Things}
\newacronym{AI}{AI}{artificial intelligence}
\newacronym{PDF}{PDF}{probability density function}
\newacronym{CF}{CF}{compress-and-forward}
\newacronym{DF}{DF}{decode-and-forward}
\newacronym{AF}{AF}{amplify-and-forward}
\newacronym{LF}{LF}{lossy-forward}
\newacronym{CEO}{CEO}{chief executive officer}
\newacronym{NOMA}{NOMA}{non-orthogonal multiple access}
\newacronym{SF}{SF}{semantic-forward}
\newacronym{LDPC}{LDPC}{low-density parity-check}
\newacronym{RD}{R-D}{relay-destination}
\newacronym{SR}{S-R}{source-relay}
\newacronym{SD}{S-D}{source-destination}
\newacronym{ED}{ED}{Euclidean distance}
\newacronym{SIC}{SemantIC}{semantic interference cancellation}
\newacronym{PSNR}{PSNR}{peak signal-to-noise ratio}
\newacronym{BER}{BER}{bit error rate}
\newacronym{6G}{6G}{sixth-generation}
\newacronym{LLR}{LLR}{log-likelihood ratio}

\begin{document}
\title{SemantIC: Semantic Interference Cancellation \\ Towards 6G Wireless Communications}
\newcommand{\authorsize}{}
\author{\authorsize
Wensheng Lin,~\IEEEmembership{\authorsize Member,~IEEE},
Yuna Yan,
Lixin Li,~\IEEEmembership{\authorsize Member,~IEEE}, 

Zhu Han,~\IEEEmembership{\authorsize Fellow,~IEEE}
and Tad Matsumoto,~\IEEEmembership{\authorsize Life Fellow,~IEEE}
 
\thanks{
This letter has been accepted for publication in IEEE Communications Letters with DOI: \href{https://doi.org/10.1109/LCOMM.2024.3412973}{10.1109/LCOMM.2024.3412973}.

Corresponding author: Lixin Li.

W. Lin, Y. Yan and L. Li are with the School of Electronics and Information, Northwestern Polytechnical University, Xi'an, Shaanxi 710129, China (e-mail: linwest@nwpu.edu.cn; yanyuna@mail.nwpu.edu.cn; lilixin@nwpu.edu.cn).

Z. Han is with the Department of Electrical and Computer Engineering at the University of Houston, Houston, TX 77004 USA, and also with the Department of Computer Science and Engineering, Kyung Hee University, Seoul, South Korea, 446-701 (e-mail: hanzhu22@gmail.com). 

T. Matsumoto is an Invited Professor at IMT-Atlantique, France.  He is also Professor Emeritus of Japan Advanced Institute of Science Technology, Ishikawa 923-1292, Japan, and University of Oulu, Finland (e-mail: matumoto@jaist.ac.jp).
}
}
\markboth{}{}
\maketitle

\begin{abstract}
This letter proposes a novel anti-interference technique, semantic interference cancellation (SemantIC), for enhancing information quality towards the sixth-generation (6G) wireless networks.
SemantIC only requires the receiver to concatenate the channel decoder with a semantic auto-encoder.
This constructs a turbo loop which iteratively and alternately eliminates noise in the signal domain and the semantic domain.
From the viewpoint of network information theory, the neural network of the semantic auto-encoder stores side information by training, and provides side information in iterative decoding, as an implementation of the Wyner-Ziv theorem.
Simulation results verify the performance improvement by SemantIC without extra channel resource cost.

\end{abstract}

\section{Introduction}
How to implement the Wyner-Ziv theorem \cite{wyner1976rate} in practical coding design? 
Academia and industry have struggled with this question for several decades, although the Wyner-Ziv theorem provides a prospect to enhance the information quality by providing side information at the decoder only.
The challenge of practical implementation is to find a method to manage side information. 

Thanks to recent development of semantic communications \cite{Yang2023Semantic}, which is a promising technological basis for the oncoming \gls{6G} wireless networks \cite{Wang20236G}, we can obtain the solution to representing side information by semantic neural networks.
In semantic communications, the semantic encoder extracts and transmits the semantic information, i.e., the feature of the original information.
Then, the receiver generates and reconstructs the original information from the semantic information by the semantic decoder \cite{Hao2024SCAI}. 
By this means, the payload transmitted through the channel can be significantly reduced.
From the viewpoint of network information theory, the semantic encoder/decoder pair learns and shares the same side information by joint training.
Hence, when separately deploying the semantic encoder and decoder, the transmitter can compress the source information according to side information, and the decoder can recover the source information with the assistance of side information.

It is noticed that the function of semantic decoder can be regarded as generative \gls{AI} \cite{Xu2023Unleashing, Xu2023Sparks}.
The diffusion model \cite{Rombach2022High} is a generative \gls{AI} method effective for achieving super-resolution, which can be also regarded as a process of eliminating noise.
Particularly, in the reverse diffusion process, noise is predicted by neural networks and subtracted from the image.
Therefore, generative \gls{AI} implicitly contains the capability for eliminating noise.
Similarly, if we train the semantic encoder and decoder by adding noise between them, the semantic encoder and decoder will also work better in denoising.

\begin{figure}[!t]
\centering 
\includegraphics[width=\figwidthT]{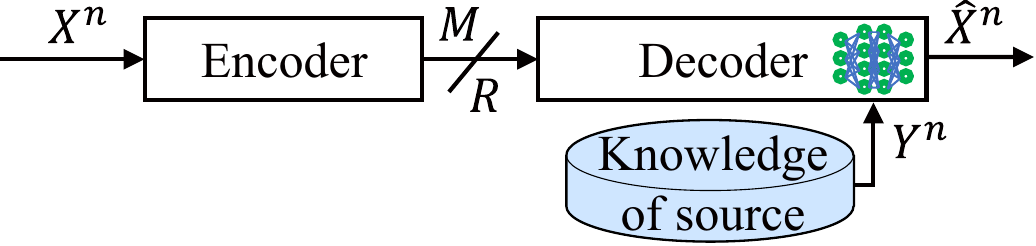}
\caption{The system model from the viewpoint of network information theory. $X^n$, $Y^n$ and $\hat{X}^n$ stand for the source information, the side information and the reconstructed information, respectively, with the sequence length of $n$.
$M$ is the codeword satisfying the coding rate $R$.}
\label{fig:WZ_model}
\end{figure}

Inspired by the Wyner-Ziv theorem and semantic communications, this letter proposes a breakthrough technique, namely \gls{SIC}, to improve the information quality by updating the receiver structure only.
The theoretical model of \gls{SIC} is exactly the Wyner-Ziv model as illustrated in \fref{fig:WZ_model}, where the side information (in practice, the knowledge of source) is learned by neural networks and stored in the form of the neural network parameters at the decoder.
The proposed \gls{SIC} technique naturally exploits the semantic information contained in the data, and hence it well matches the feature of 6G which focuses more on the semantic information of data.

In practical implementation, the \gls{SIC} receiver concatenates a pair of semantic encoder/decoder with the channel decoder. 
The semantic encoder extracts the semantic information of the channel decoder output.
In this way, the noise and interference can be eliminated to a certain extent when extracting information features.
From the viewpoint of machine learning, the concatenated semantic encoder/decoder pair constructs an auto-encoder \cite{Vincent2010Stacked, Luo2022Autoencoder}, which has also been verified to be able to eliminate noise \cite{Du2017Stacked}.
Subsequently, based on the denoised information features, the semantic decoder reconstructs the original information to input into the channel decoder as the \emph{a priori} information for the next round of iterative decoding.

Note that the channel decoder concatenated with the semantic auto-encoder composes a turbo loop.
Owing to the turbo principle \cite{berrou1996near}, the decoding gain is alternately enhanced by the channel decoder and the semantic auto-encoder. 
Furthermore, since the side information is introduced into the turbo loop by the semantic auto-encoder, the final decoding output contains more information and hence achieves a higher quality.
The contributions of this letter is summarized as follows.
\begin{itemize}
\item We propose a novel anti-interference technique, i.e., \gls{SIC}, which can improve the recovered information quality without extra cost of channel resources.
\item We design an iterative decoding algorithm for image transmissions with \gls{SIC} by constructing a turbo loop concatenating the channel decoder and the semantic auto-encoder, which alternately eliminates noise in the signal domain and the semantic domain.
\item Simulation results show that the system with \gls{SIC} has a better performance in terms of \gls{BER}, \gls{ED}, and \gls{PSNR}, 
even with a very light neural network (42.1 kB).
\end{itemize}

\section{Semantic Interference Cancellation Principle}
\subsection{Theoretical Basis}
The theoretical basis of \gls{SIC} is the Wyner-Ziv theorem.
As illustrated in \fref{fig:WZ_model}, the Wyner-Ziv system is required to transmit a source sequence $X^n$ under the rate constraint $R$.
Therefore, an encoder compresses the source sequence $X^n$ into a codeword $M$ before transmission.
Then, at the decoder, the noncausal side information sequence $Y^n$ is available to help better reconstruct the source sequence.
The recovery $\hat{X}^n$ may be lossy if both the channel capacity and the side information is insufficient.
Mathematically, the link rate $R$ should satisfy the rate-distortion function given a distortion requirement $D$, well known as the Wyner-Ziv theorem: 
\begin{align}
R(D) \ge I(X;U|Y), \label{eq:WZ}
\end{align}
where $U$ represents the compress information of $X$.

In practical systems, the mutual information gain provided by the knowledge base can be mathematically evaluated as 
\begin{align}
I_{gain}=I(X;\hat{X}_{S})-I(X;\hat{X}_C), \label{eq:I_gain}
\end{align}
where $\hat{X}_{S}$ is the information recovered with the help of the semantic auto-encoder, and $\hat{X}_C$ is the information independently recovered by the channel decoder only.
Both $I(X;\hat{X}_S)$ and $I(X;\hat{X}_{C})$ can be calculated from the mutual information between the transmitted and the recovered information.

Although Wyner and Ziv have presented a coding scheme in the proof of the rate-distortion function in \eqref{eq:WZ}, their proposed algorithm requires the sequence length tends to $\infty$.
Moreover, since their coding scheme utilizes a random codebook, it requires a huge memory to store the codebook, and the encoding/decoding process is of very high time-complexity due to exhaustive search over the random codebook.

To implement the Wyner-Ziv theorem in practical systems, we have to solve many tricky problems including how to obtain, store and represent side information, as well as how to exploit side information in decoding without using a random codebook.
We find that neural networks can automatically learn the knowledge about the source by training, which is essentially the same as obtaining, storing and representing side information.
It is also noticed that the side information stored in the form of neural network parameters is noncausal, which satisfies the condition of the Wyner-Ziv theorem.
Now, we need to find the last piece of the puzzle, i.e., decoding with the side information stored in neural networks.

\subsection{Decoder Structure}

\begin{figure}[!t]
\centering \includegraphics[width=\figwidthS]{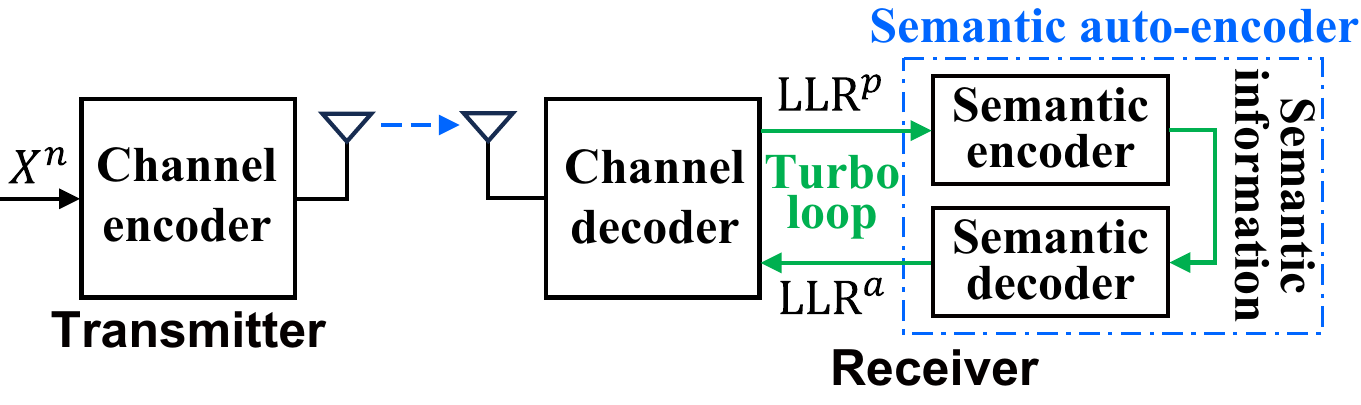}
\caption{The structure of SemantIC system.}
\label{fig:DEC}
\end{figure}

To provide side information by neural networks for decoding, we need to design the decoder structure.
Therefore, we propose the structure of the \gls{SIC} system in \fref{fig:DEC}, where the transmitter has a general structure without constraint.
At the receiver, the channel decoder $\mathcal{F}_C(\cdot)$ processes the \emph{a priori} \gls{LLR} to output the \emph{a posteriori} \gls{LLR} as
\begin{align}
\mathrm{LLR}^p = \mathcal{F}_C(\mathrm{LLR}^a).\label{eq:F_C}
\end{align}
The \gls{SIC} decoder contains two key elements as follows.

\subsubsection{Auto-Encoder}
The auto-encoder is a neural network having the function to extract information features and then recover information from features.
By this method, the semantic noise is filtered by the neural network.
In the SemantIC system, the receiver utilizes a semantic auto-encoder, consists of a pair of semantic encoder/decoder, to eliminate the interference and noise contained in the channel decoder output. 
This process is referred to as denoising in the semantic domain.

\subsubsection{Turbo Loop}
To enhance the decoding gain, the structure of iterative decoding is necessary.
Therefore, we connect the output of the semantic auto-encoder $\mathcal{F}_S(\cdot)$ back to the channel decoder $\mathcal{F}_C(\cdot)$  as the \emph{a priori} information.
\begin{align}
\mathrm{LLR}^a = \mathcal{F}_S(\mathrm{LLR}^p) \label{eq:F_S}.
\end{align}
The structure of \eqref{eq:F_C} and \eqref{eq:F_S} composes a turbo loop for eliminating noise and interference in the signal domain and the semantic domain alternately.
The semantic auto-encoder performs semantic denoising from the input $\mathrm{LLR}^p$, and then output $\mathrm{LLR}^a$ to the channel decoder for denoising of signals in the following iteration round of decoding.

Based on the semantic auto-encoder and the turbo loop, the \gls{SIC} system can further improve the reconstructed information quality without any change outside the receiver.
The turbo loop reduces noise at the cost of introducing extra latency in each round of iterations by the semantic auto-encoder depending on the complexity of the neural networks.

\section{Decoder Implementation Design}
In order to evaluate the performance of the \gls{SIC} system, we design a decoding algorithm for image transmission simulations.
Since the semantic encoder/decoder requires a dataset for training, we select the CIFAR-10 dataset \cite{CIFAR10} for performance evaluation.
The CIFAR-10 dataset comprises a collection of 60,000 pictures, consisting of 50,000 training images and an additional set of 10,000 images specifically designed for testing purposes. 
The image size is scaled to $3 \times 96 \times 96$, i.e., 3 color channels, and 96 pixels in both width and height.

\begin{figure}[!t]
\centering \includegraphics[width=0.9\linewidth]{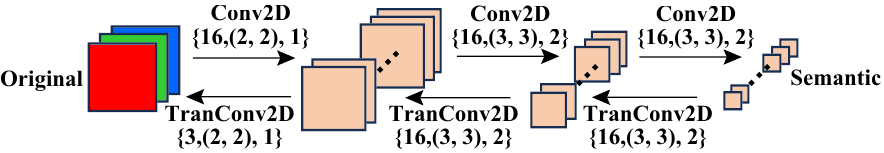}
\caption{A simple structure of semantic encoder/decoder.}
\label{fig:ENC_DEC}
\end{figure}

As depicted in \fref{fig:ENC_DEC}, we concatenate two-dimensional (2D) convolutional layers (Conv2Ds) as the semantic encoder, and 2D transpose convolutional layers (TranConv2Ds) are concatenated to be the semantic decoder.
Let $\{n_F,(w_K,h_K),s\}$ specify the parameters of Conv2D and TranConv2D, with $n_F$ denoting the number of filters, $(w_K,h_K)$ denoting the kernel size, and $s$ denoting the stride.
The parameters are configured to gradually extract the semantic features with a rate approximate to $0.3$, i.e., the output size is $16 \times 23 \times 23$.
As will be shown in the simulation results, even with such simple semantic encoder/decoder structure, the \gls{SIC} decoder is effective with respect to improving information quality.

Our model is implemented using Pytorch, with training conducted on the NVIDIA GeForce RTX 3060 Ti by the Adam optimizer \cite{kingma2014adam}.
Compared to the 163 MB size of the CIFAR-10 dataset, the file for storing the semantic auto-encoder parameter only has a size of 42.1 kB.

\Gls{LDPC} codes \cite{gallager1962low} are selected as the channel code, with the codeword length being 900.
The number of parity-check equations including a certain bit is set at 2, and the number of bits in the same parity-check equation is set at 3.
The maximum iteration round is set at 7.
The information of each image is divide into groups to match the LDPC codeword length for transmission. 
Since the LDPC codes work for binary sequences, the image information in the form of pixels is quantized into binary sequences before inputting into the channel decoder.
After channel decoding, the output should also be dequantized back to pixels.

An example source code for the implementation of the proposed \gls{SIC} system is presented in \cite{SemantICgithub}.

\begin{figure}[!t]
\centering 
\subfigure[BER.]{
\includegraphics[width=\figwidthS]{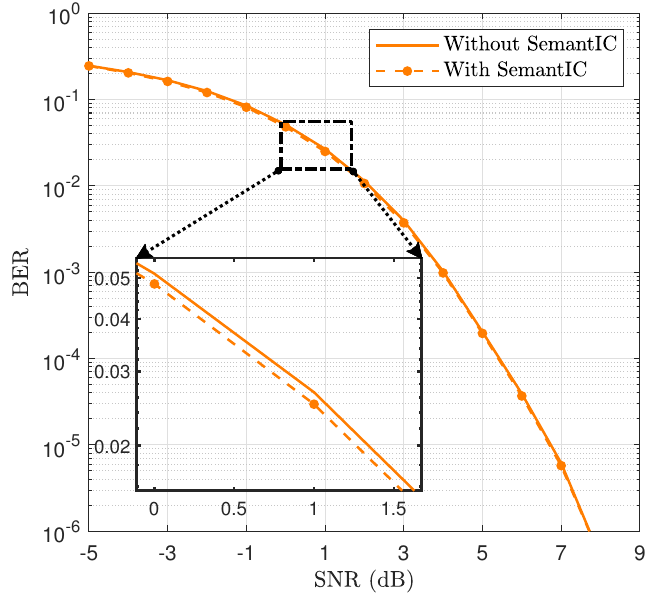}
\label{fig:BER}
}
\subfigure[\Gls{ED} and PSNR.]{
\includegraphics[width=3.1in]{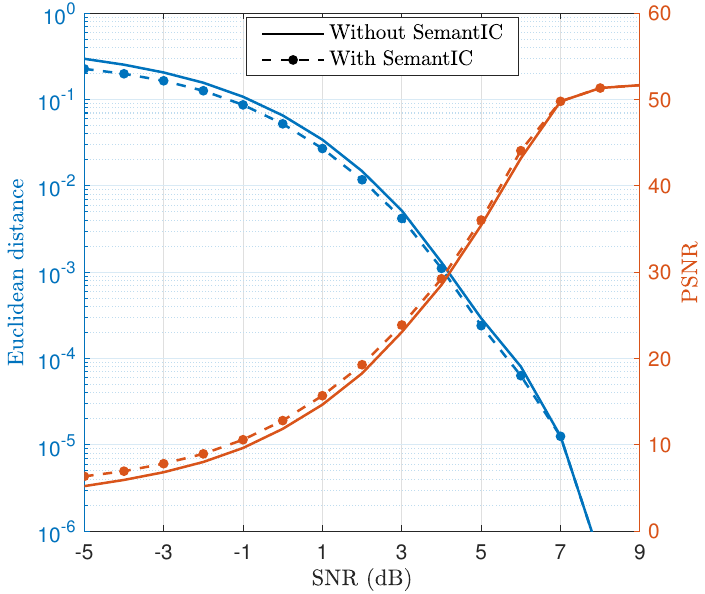}
\label{fig:ED_PSNR}
}
\caption{The impact of SNR on BER, \gls{ED}, and PSNR.}
\label{fig:vsSNR}
\end{figure}

\section{Simulation Results}
In this section, we conduct a comparative analysis between the system with and without SemantIC by simulations. 
We employ the \gls{BER}, the \gls{ED}, and the \gls{PSNR} as evaluation metrics to quantify the distortion between the original images $\boldsymbol{X}$ and the reconstructed images $\boldsymbol{\hat{X}}$. 
Additionally, we set the learning rate to 0.003, the batch size to 64, and the maximum training epoch to 200.

\subsubsection{Anti-Interference Performance} As shown in \fref{fig:vsSNR}, we demonstrate the relationship between the BER, \gls{ED}, PSNR and SNR of the proposed method in an \gls{AWGN} channel. 
It is easy to observe from the figure that the SemantIC system outperforms the system without SemantIC in terms of lower BER and \gls{ED} as well as higher PSNR at different SNR values.
It is observed that the BER performance gain seems less visible than \gls{ED}. This is because the loss function for training the semantic auto-encoder is designed based on \gls{ED} instead of BER.
Although the BER performance seems to be very small, this performance gain is obtained without extra channel resource cost, and the \gls{SIC} neural network has a very simple structure and with a very small size of parameters.
The system is expected to have a better anti-interference performance, if the neural network is sophisticatedly designed, e.g., introducing the attention mechanism to enlarge receptive field, and/or utilizing the diffusion model to predict the noise.

Particularly, SemantIC performs better in low SNR scenarios.
This can be attributed to the fact that higher-level semantic features are more resilient to channel noise and fluctuations compared to lower-level semantic features. 
By extracting high-level semantic features through the semantic encoder/decoder, SemantIC aids the channel decoder in recovering lost semantic information during the physical channel transmission process, thereby enhancing system robustness.
Another factor for lower gain by \gls{SIC} is that LDPC codes have a strong capability for enhancing decoding gain.
Hence, when SNR is large, LDPC codes have already significantly eliminated noise and interference.
Especially when SNR$\ge 8$ dB, the BER by channel decoding only has reduced to $0$ at the first round of iteration. 
Therefore, no gain can be obtained for SNR$\ge 8$ dB.

\begin{figure}[!t]
\centering 
\subfigure[BER.]{
\includegraphics[width=\figwidthM]{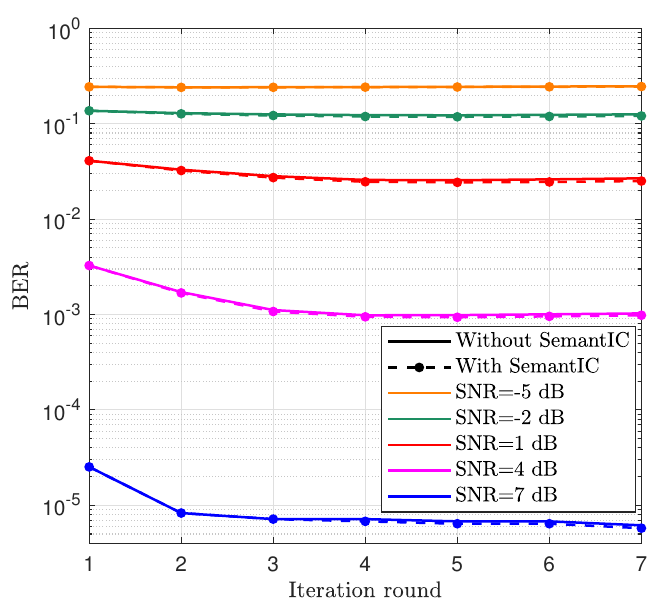}
\label{fig:BER_converge}
}
\subfigure[\gls{ED}.]{
\includegraphics[width=\figwidthM]{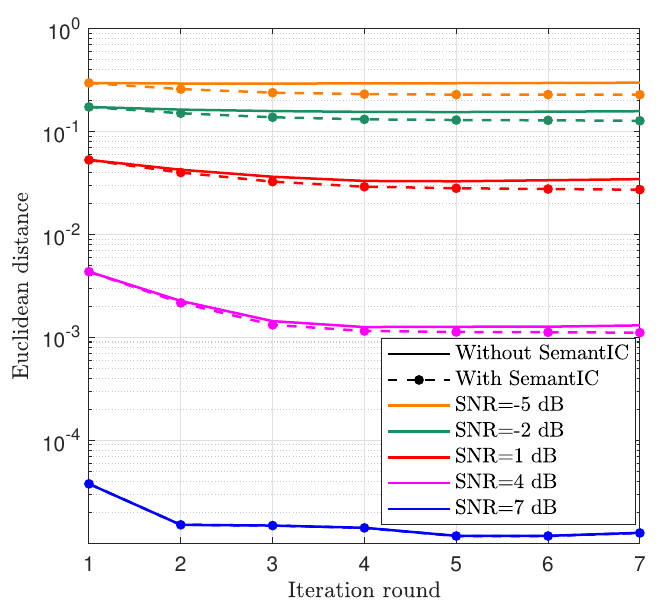}
\label{fig:ED_converge}
}
\subfigure[PSNR.]{
	\includegraphics[width=\figwidthM]{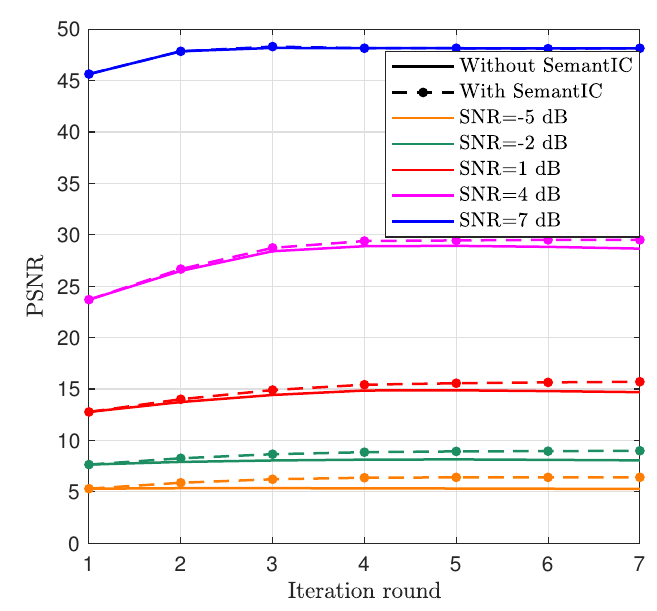}
	\label{fig:PSNR_converge}
}
\caption{Convergence behaviour for BER, \gls{ED}, and PSNR.}
\label{fig:converge}
\end{figure}

\subsubsection{Iterative Gain} 
\fref{fig:converge} demonstrates the performance gain achieved through iterative decoding without extra channel resource cost. 
It is evident that as the number of iterations increases, SemantIC consistently outperforms the method without SemantIC, and the performance gap gradually becomes significant. 
This is because during the iterative decoding process, the semantic auto-encoder can filter out noise and interference to a certain extent and provide additional semantic information through multiple interactions. 
This observation verifies the effectiveness of the turbo loop in the \gls{SIC} decoder.
Consequently, it effectively reduces error propagation and enhances overall decoding performance.
Moreover, the performance converges fast at the fourth round of iteration. 
This is due to the strong error correcting capability of LDPC codes.
Hence, practical systems can adopt early stopping strategy to balance computational complexity with latency.
Interestingly, as SNR becomes larger, the BER performance gain increases, while the performance gain of the \gls{ED} and PSNR decreases.


\begin{figure}[!th]
\centering 
\subfigure[Lossless images.]{
\includegraphics[width=1.3in]{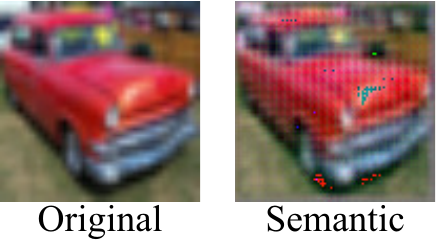}
\label{fig:origin_semantic}
}
\subfigure[SNR$=-5$ dB.]{
\includegraphics[width=\figwidthW]{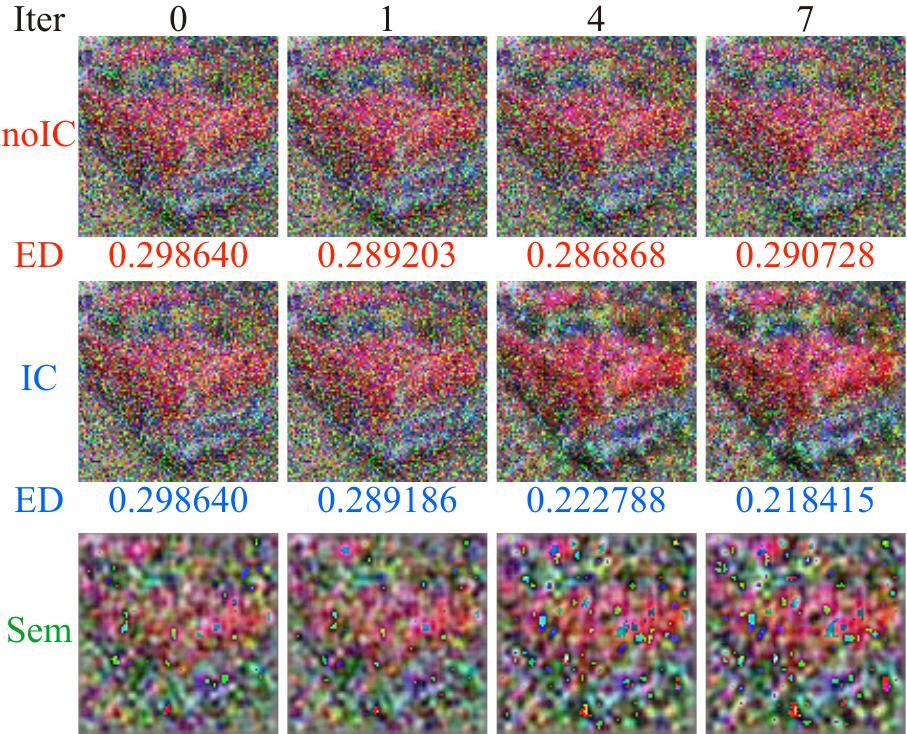}
\label{fig:snr-5}
}
\subfigure[SNR$=0$ dB.]{
\includegraphics[width=\figwidthW]{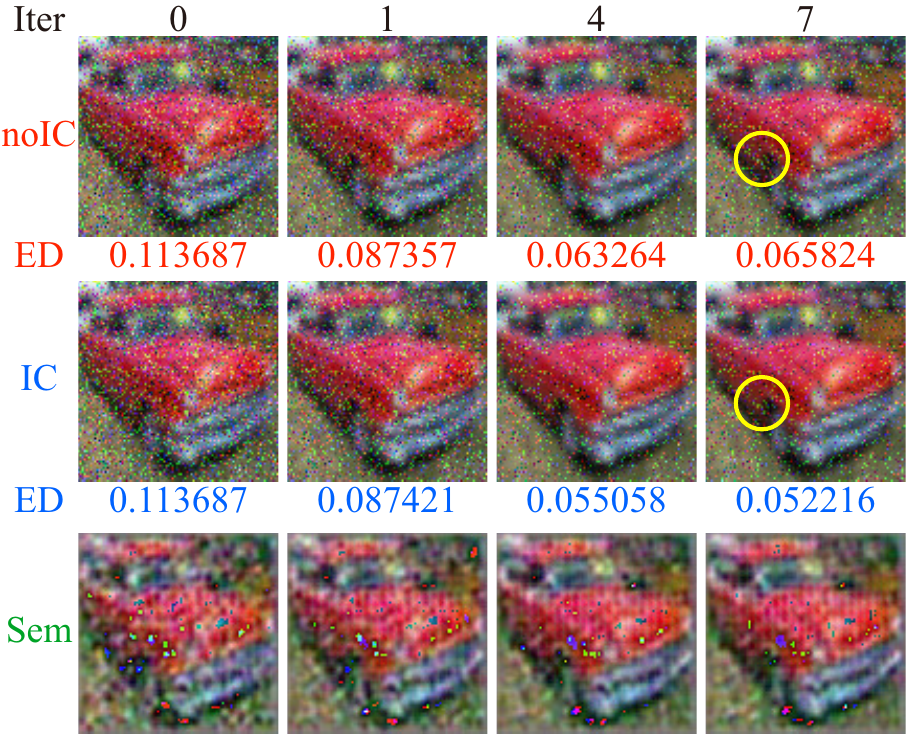}
\label{fig:snr0}
}
\subfigure[SNR$=5$ dB.]{
\includegraphics[width=\figwidthW]{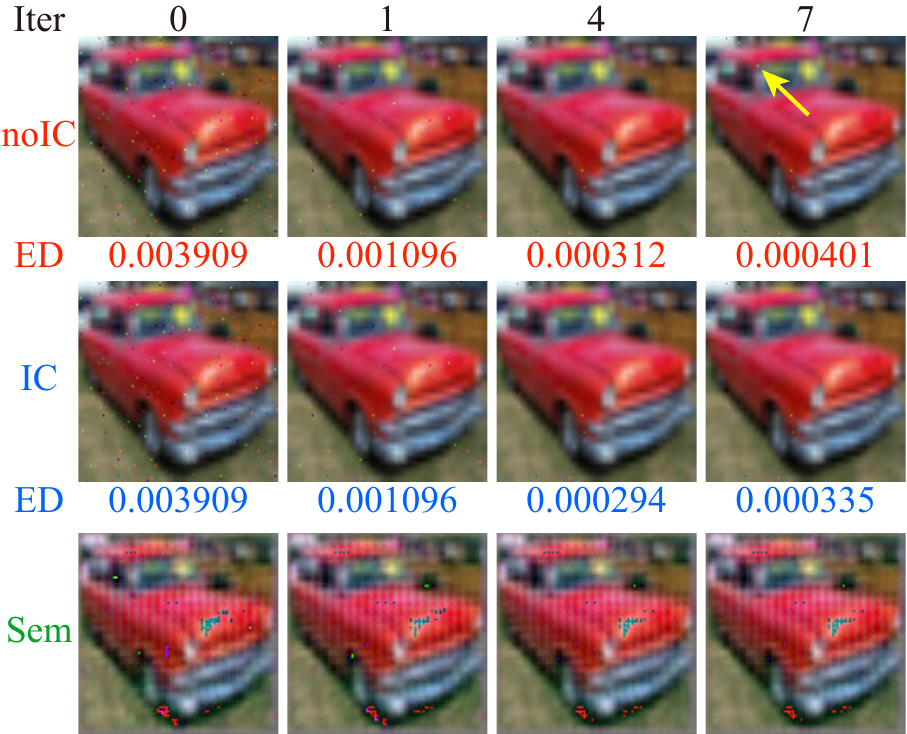}
\label{fig:snr5}
}
\caption{Comparisons of the reconstructed image quality.}
\label{fig:image_quality}
\end{figure}

\subsubsection{Visual Effect} 
For the comparison of visual effect, we utilize the original image presented in \fref{fig:origin_semantic} as an example, and the semantic image is obtained by inputting the original image to the semantic encoder concatenated with the semantic decoder.
Fig. \ref{fig:image_quality} visualizes the results of image reconstruction with and without SemantIC, subjectively verifying the fidelity of semantic information. 
In \fref{fig:image_quality}, Iter, IC/noIC, Sem, ED represent the iteration rounds, the image with/without interference cancellation, the image reconstructed by the semantic decoder, and \gls{ED}.

In Fig. \ref{fig:snr-5}, the method without SemantIC fails to clearly depict the shape of the car window, whereas our approach successfully recovers some details of the window from the blurred original image. 
In Fig. \ref{fig:snr0}, SemantIC enhances both the visual quality and semantic information of the image. 
Traditional algorithms often focus solely on visual quality, neglecting the preservation of semantic information. 
However, SemantIC simultaneously considers both image quality and semantic information, rendering objects and scenes in the image more realistic, recognizable, and comprehensible.
If we zoom-in and compare the image of IC/noIC with Iter$=7$, we can see that hot pixels within the yellow circle is less for the IC case. 
The semantic image also gradually becomes clearer as the iteration round increases.
In Fig. \ref{fig:snr5}, as channel conditions improve, the reconstructed image becomes highly realistic, making it difficult to discern from the original image. 
Although the image quality is already very high without SemantIC, our approach can still further eliminate some noise.
For example, if we zoom-in the image with Iter$=7$, we can see that one hot pixel pointed by the yellow arrow is corrected.

\section{Conclusion}
This letter has proposed a breakthrough technique, referred to as \gls{SIC}, for anti-interference towards the 6G wireless communications.
The \gls{SIC} decoder contains a turbo loop consists of concatenated channel decoder and semantic auto-encoder, which eliminate the noise and interference alternately in signal domain and semantic domain.
We have also designed a decoding algorithm and implemented the \gls{SIC} system for image transmissions.
The simulation results indicate that the \gls{SIC} system can enhance the reconstructed information quality without extra cost of channel resources, although we only utilize a very simple neural network structure.
Even with a very low SNR, the image recovered by \gls{SIC} shows a clearer object shape than that without \gls{SIC}.

\bibliographystyle{IEEEtran}
\bibliography{myreference}

\end{document}